\documentclass[epj]{svjour}
\usepackage{graphics}
\begin{document}
\title{$B_s$ Properties at the Tevatron}
\author{Guillelmo G\'omez-Ceballos\thanks{\email{ceballos@fnal.gov}}
}                     
%
%
\institute{Instituto De F\'{\i}sica de Cantabria, Avda. de los Castros s/n, 39002 Santander, Cantabria, Spain}
%
\date{Received: 08/05/2005}
%
\abstract{
The Tevatron collider at Fermilab provides a very rich environment for the study 
$B_s$ mesons. In this paper we will show a few selected topics from the CDF and 
D$\oslash$ collaborations, giving special attention to the $B_s$ Mixing 
analyses.
\PACS{
      {Tevatron}, {$B_s$}, {Mixing}, {$\Delta\Gamma_s$/$\Gamma_s$}, {$\Delta m_s$} 
     } 
} 
\maketitle
\section{Introduction}
\label{sec:intro}
The Tevatron collider at Fermilab, operating at $\sqrt{s}$ =1.96 $TeV$, has a huge $b$ 
production rate which is 3 orders of magnitude higher than the production rate at $e^+e^-$ 
colliders running on the $\Upsilon(4S)$ resonance. Among the produced $B$ particles there 
are as well heavy and excited states which are currently uniquely accessible at the Tevatron, 
such as for example $B_s$, $B_c$, $\Lambda_b$, $\theta_b$, $B^{**}$ or $B^{**}_s$. Dedicated 
triggers are able to pick 1 $B$ event out of 1000 QCD events by selecting leptons and/or 
events with displaced vertices already on hardware level.

The aim of the $B$ Physics program of the Tevatron experiments CDF and D$\oslash$ is to provide 
constraint to the CKM matrix which takes advantage of the unique features of a hadron 
collider. Several topics related to $B_s$ mesons were discussed by other speakers in the 
conference, therefore we will focus this paper in three flaship analyses: 
$B_s \to h^{+} h^{'-}$, $\Delta\Gamma_s/\Gamma_s$ and 
$\Delta m_s$~\cite{cdfpublic,d0public}.

Both the CDF and the D$\oslash$ detector are symmetric multi-purpose detectors having both 
silicon vertex detectors, high resolution tracking in a magnetic field and lepton 
identification~\cite{cdf,d0}. CDF is for the first time in an hadronic environment able to trigger
 on hardware level on large track impact parameters which indicates displaced vertices. 
Thus it is very powerful in fully hadronic $B$ modes. 

\section{$B_{s(d)} \to h^{+} h^{'-}$ Decays}
\label{sec:bhh}
\begin{figure}
\begin{center}
\resizebox{0.50\textwidth}{0.20\textheight}{%
  \includegraphics{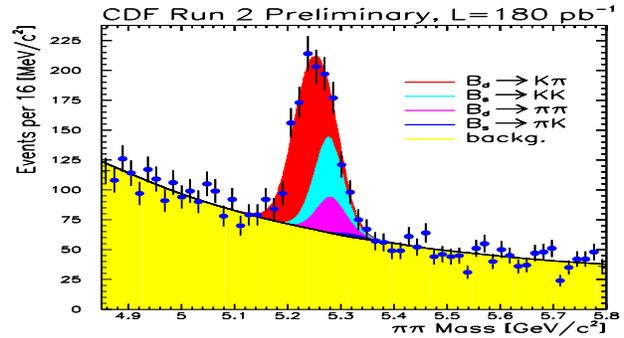}
}
\caption{$\pi \pi$ invariant mass distribution of $B_{s(d)} \to h^{+}h^{'-}$ candidates.}
\label{fig:mpipi_color2}       
\end{center}
\end{figure}

\begin{figure}
\begin{center}
\resizebox{0.50\textwidth}{0.20\textheight}{%
  \includegraphics{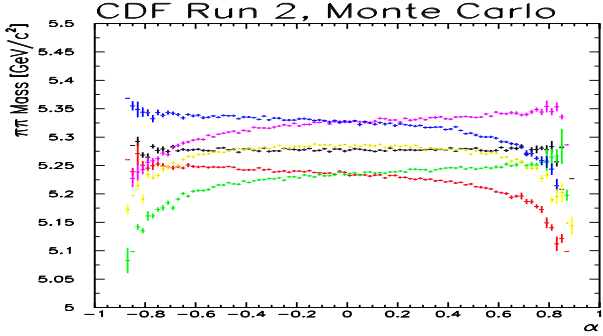}
}
\caption{Monte Carlo distribution of $M_{\pi\pi}$ versus $(1-\frac{p_1}{p_2})*q_1$ 
for different $B_{s(d)} \to h^{+}h^{'-}$ channels.}
\label{fig:prof_tutti}       
\end{center}
\end{figure}

Using the new trigger on displaced tracks, CDF has collected several hundred events of charmless
$B_d$ and $B_s$ decays in two tracks. The invariant mass spectrum of the $B_{s(d)} \to h^{+}
h^{'-}$ candidates with pion mass assignment for both tracks is shown in 
Fig.~\ref{fig:mpipi_color2}. A clear peak is seen, but with a width much larger than the 
intrinsic CDF resolution due to the overlap of four different channels under the peak: 
$B_d \rightarrow K^+\pi^-$, $B_s \rightarrow K^+K^-$, $B_d \rightarrow \pi^+ \pi^-$ and 
$B_s \rightarrow \pi^+ K^-$. One of the goals of CDF is to measure time-dependent decay CP
asymmetries in flavor-tagged sample of $B_s \rightarrow K^+K^-$ and
$B_d \rightarrow \pi^+ \pi^-$ decays. The first step has been to disantangle the different
contributions. To do that a couple of variables has been combined in an unbinned maximum
likelihood fit in addition to the reconstructed mass. The first variable is the dE/dX 
information, which has a separation power between kaons and pions of about 1.4$\sigma$. The other
variable is the kinematic charge correlation between the invariant mass $M_{\pi\pi}$ and the
signed momentum imbalance between the two tracks, $\alpha = (1-\frac{p_1}{p_2})*q_1$, where $p_1$
($p_2$) is the scalar momentum of the track with the smaller (larger) momentum and $q_1$ is the
charge of the track with smaller momentum. The distribution from Monte Carlo simulation of 
$M_{\pi\pi}$ versus $\alpha$ is shown in Fig.~\ref{fig:prof_tutti}.

With this, we obtain the first observation of $B_s \rightarrow K^+K^-$:

\begin{center}
$\frac{f_s~R(B_s \rightarrow K^+K^-)}{f_d~BR(B_d\rightarrow K\pi)} =  
0.46\pm 0.08(stat.)\pm 0.07(syst.)$,
\end{center}

and a big improvement in the limit on $B_s \rightarrow K^+\pi^-$:

\begin{center}
$BR(B_s \rightarrow K\pi) < 0.08*BR(B_d\rightarrow K\pi)*(f_s/f_d)$\\
@90\% C.L.
\end{center}

In the $B_d$ sector we obtain:

\begin{center}
$A_{CP}(B_d \rightarrow K\pi) = 
\frac{N(B_d \rightarrow K^+\pi^-) - N(\overline{B}_d \rightarrow K^-\pi^+)}
{N(B_d \rightarrow K^+\pi^-) + N(\overline{B}_d \rightarrow K^-\pi^+)} 
= -0.022\pm0.078(stat.)\pm0.012(syst.)$,
\end{center}

being this result perfectly compatible with $B$ factories. It is important to
notice that $A_{CP}$ systematics are at the level of Babar and Belle 
experiments, and we expect to reach Y(4S) precision on the statistical 
uncertainty with the current sample on tape as well.

\section{$\Delta\Gamma_s/\Gamma_s$ Measurement in $B_s \rightarrow J/\Psi \phi$ Decays}
\label{sec:deltagamma}
In order to measure the decay width difference $\Delta\Gamma_s$ we need 
to disantangle the heavy and light $B_s$ mass eigentstates and measure 
their lifetimes separately. In the $B_s$ system CP violation is supposed 
to be small ($\delta \phi_s \approx 0$). Thus the heavy and light $B_s$ mass 
eigenstates directly correspond to the CP even and CP odd eigenstates.
So the separation of the $B_s$ mass eigenstates can be done by identifying 
the CP even and CP odd contributions.

Generally final states are mixtures of CP even and odd states, but for 
pseudoscalar particles where the $B_s$ decays into two vector particles 
such as the $J/\Psi$ and the $\phi$ it is possible to disantangle the CP 
even and CP odd eigenstates by an angular analysis. The decay amplitude 
decomposes into 3 linear polarization states with the amplitudes 
$A_0, A_{\parallel}$ and $A_{\perp}$ with
\begin{equation}
|A_0|^2 + |A_{\parallel}|^2 + |A_{\perp}|^2=1.
\end{equation}
$A_0$ and $A_{\parallel}$ correspond to the S and D wave and are therefore 
the CP even contribution, while $A_{\perp}$ corresponds to the P wave and 
thus to the CP odd component.

It is possible to measure the lifetimes of the heavy and light $B_s$ mass 
eigenstate, by fitting at the same time for the angular distributions 
and for the lifetimes.

A similar angular analysis has been already performed by the BABAR and 
BELLE experiments in the $B_d \rightarrow J/\Psi K^{*0}$ mode. This mode 
has as well been studied at the Tevatron as a cross check for the 
$B_s \rightarrow J/\Psi \phi$ analysis.

\begin{figure}
\begin{center}
\resizebox{0.50\textwidth}{0.20\textheight}{%
  \includegraphics{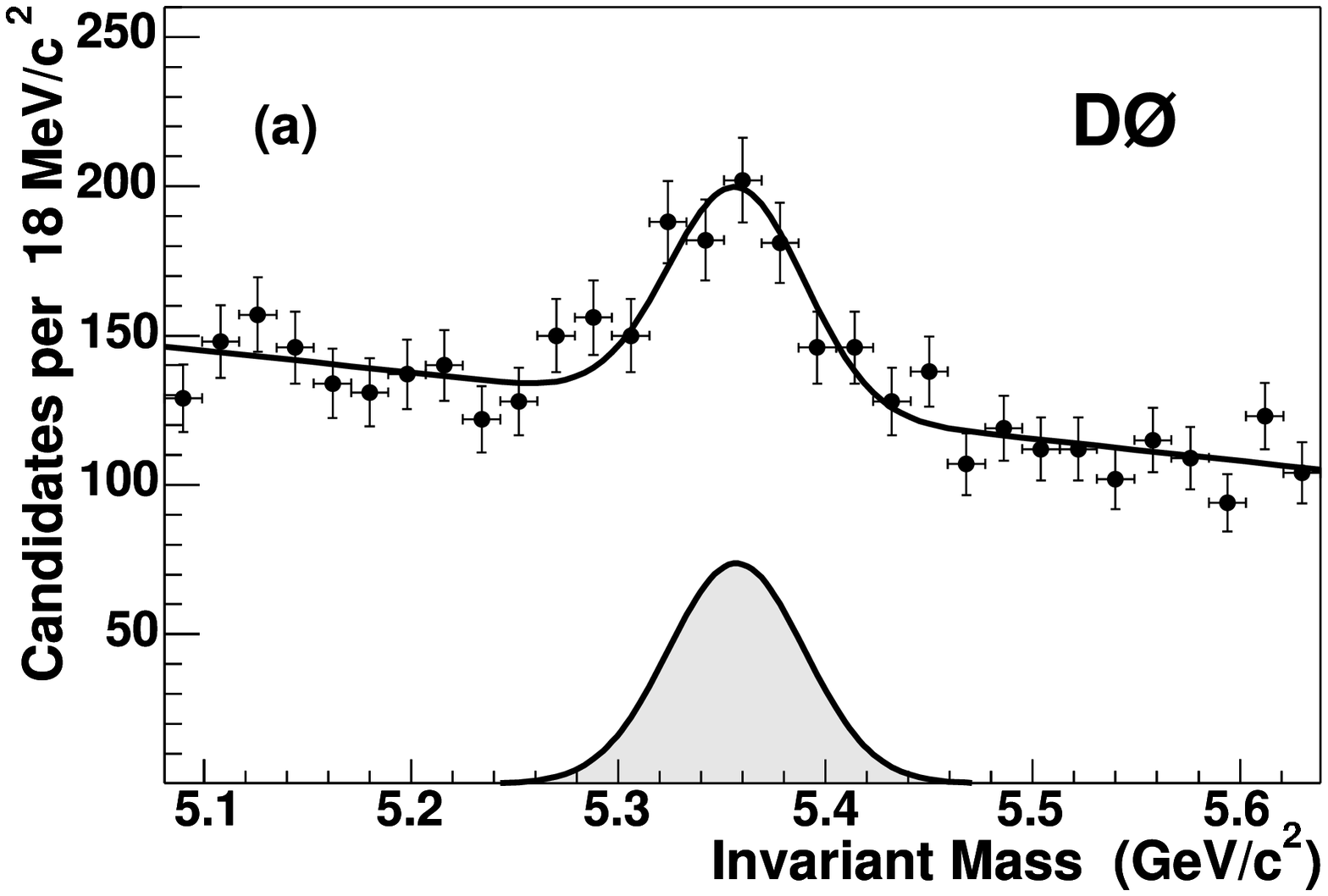}
  \includegraphics{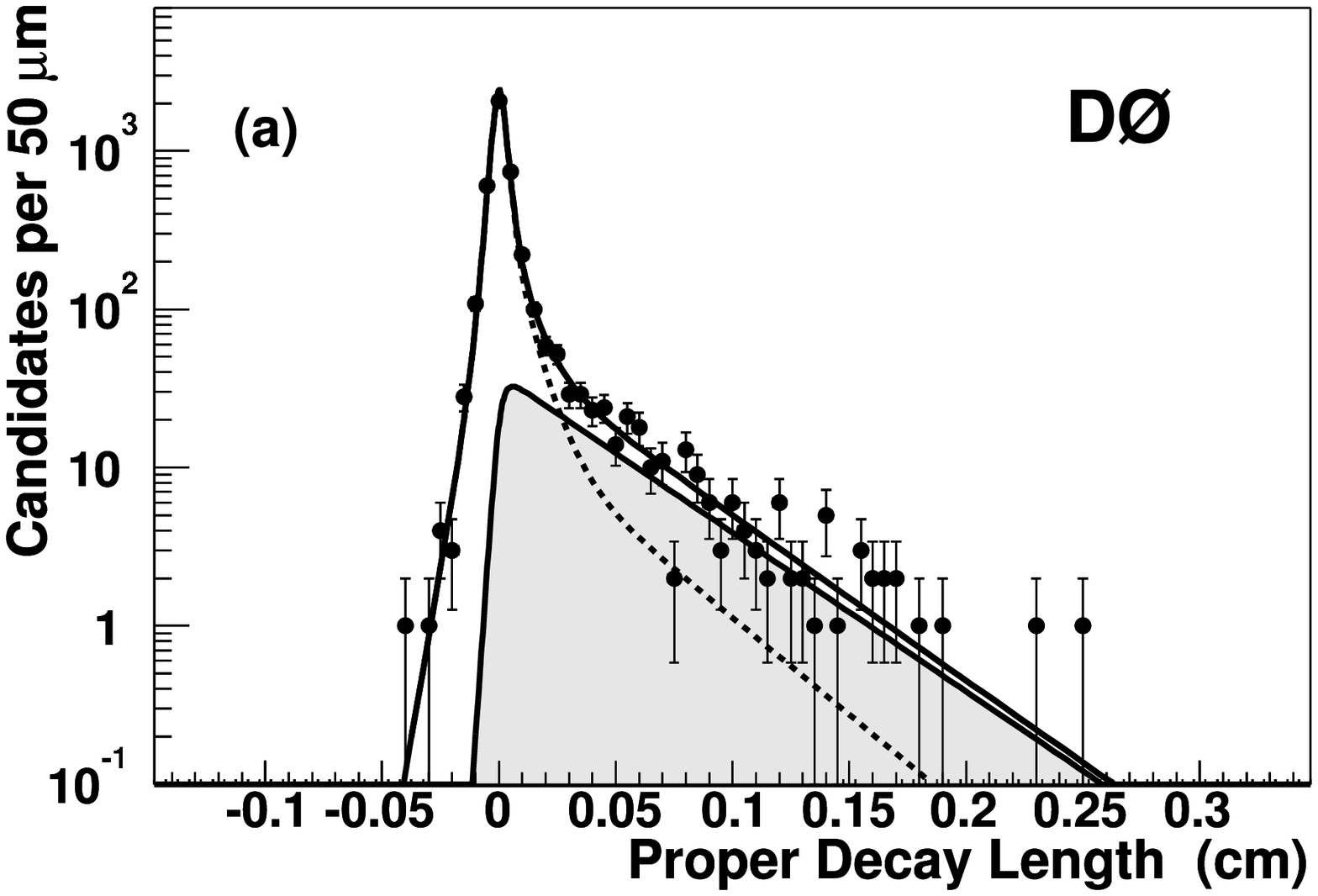}
}
\caption{Mass (left) and average lifetime (right) distributions of 
$B_s \rightarrow J/\Psi \phi$ candidates from D$\oslash$.}
\label{fig:ceballosbsjpsiphid0}       
\end{center}
\end{figure}

In order to perform this analysis first of all a 
$B_s \rightarrow J/\Psi \phi$ signal has to be established. Both experiments 
have measured the $B_s$ mass and lifetime, as shown in 
Fig.~\ref{fig:ceballosbsjpsiphid0} for the D$\oslash$ analysis, 
where the lifetime $\tau_s$ is measured with respect to $\tau_d$ from 
the topological similar decay $B_d \rightarrow J/\Psi K^{*0}$.

\begin{figure}
\begin{center}
\resizebox{0.50\textwidth}{0.22\textheight}{%
  \includegraphics{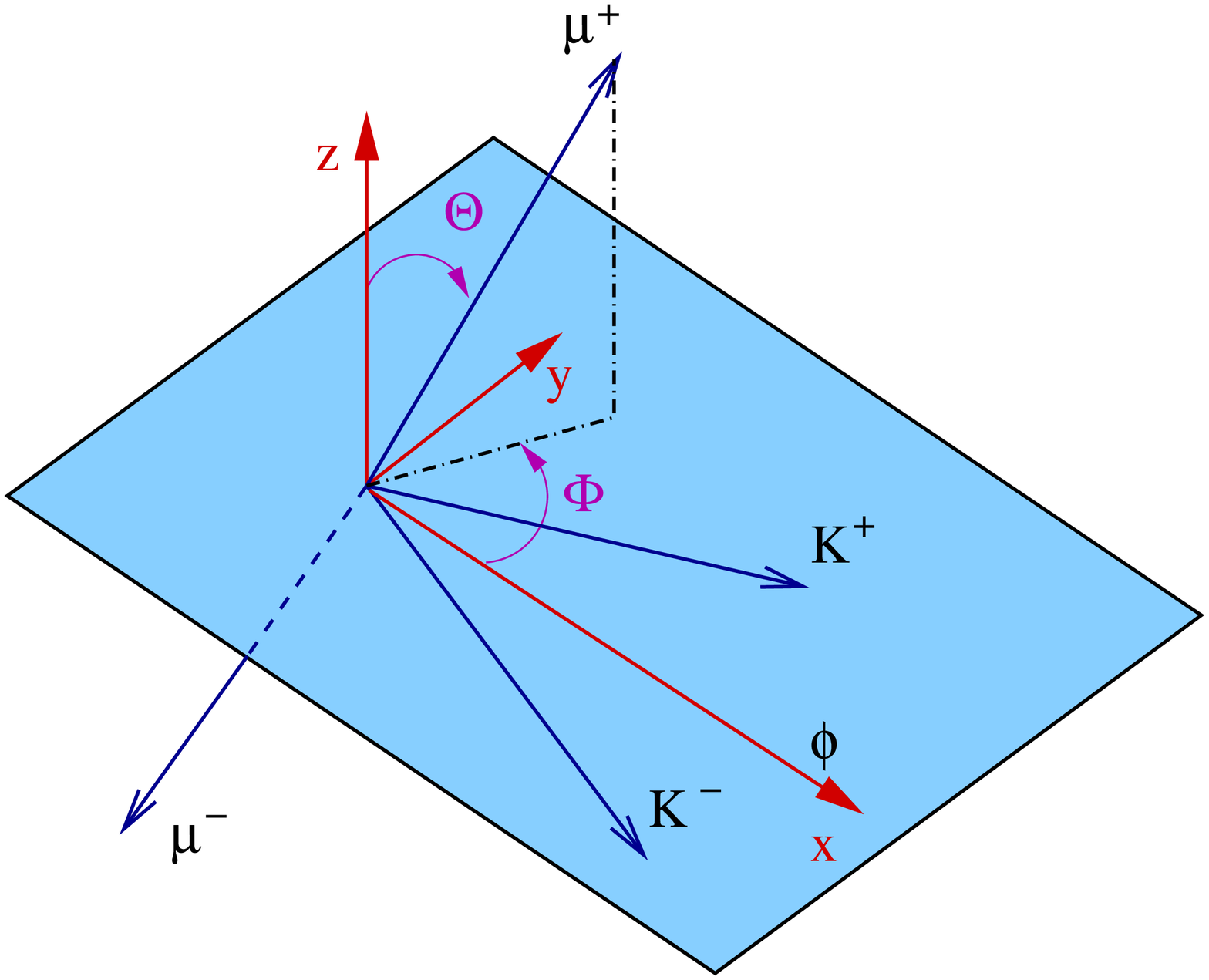}
  \includegraphics{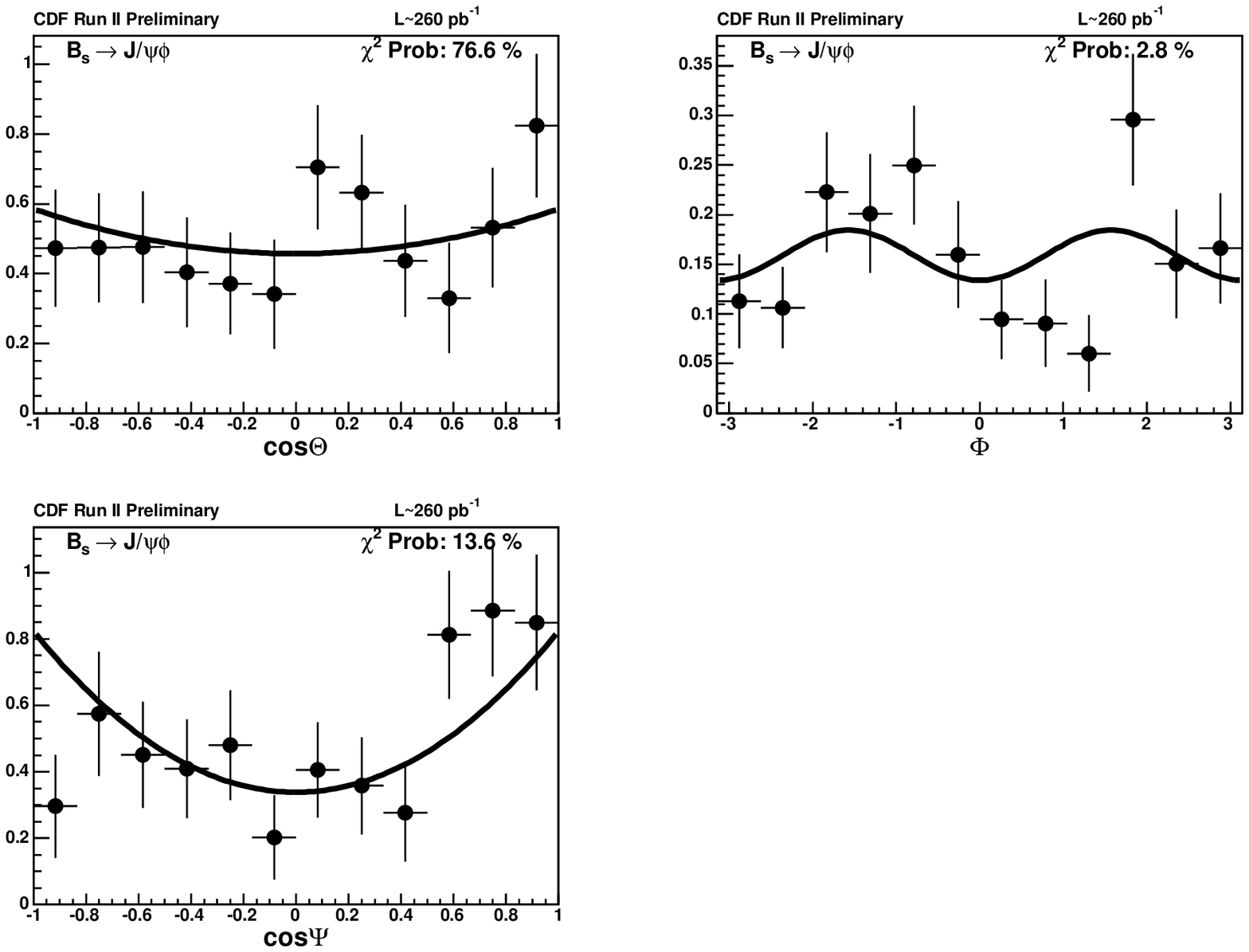}
}
\caption{Definition of the transversity frame and the transversity angles (left) and 
fit projections of the common fit of both lifetime and angular 
distributions from the CDF analysis (right).}
\label{fig:ceballos_pol}       
\end{center}
\end{figure}

The angular analysis has been performed in the transversity basis in the 
$J/\Psi$ rest-frame which is introduced in Fig.~\ref{fig:ceballos_pol}. 
The fit projections of the common fit of the both lifetimes and the angular 
distributions for the CDF analysis  and for the D$\oslash$ analysis are shown in 
Fig.~\ref{fig:ceballos_pol}.

\begin{figure}
\begin{center}
\resizebox{0.50\textwidth}{0.20\textheight}{%
  \includegraphics{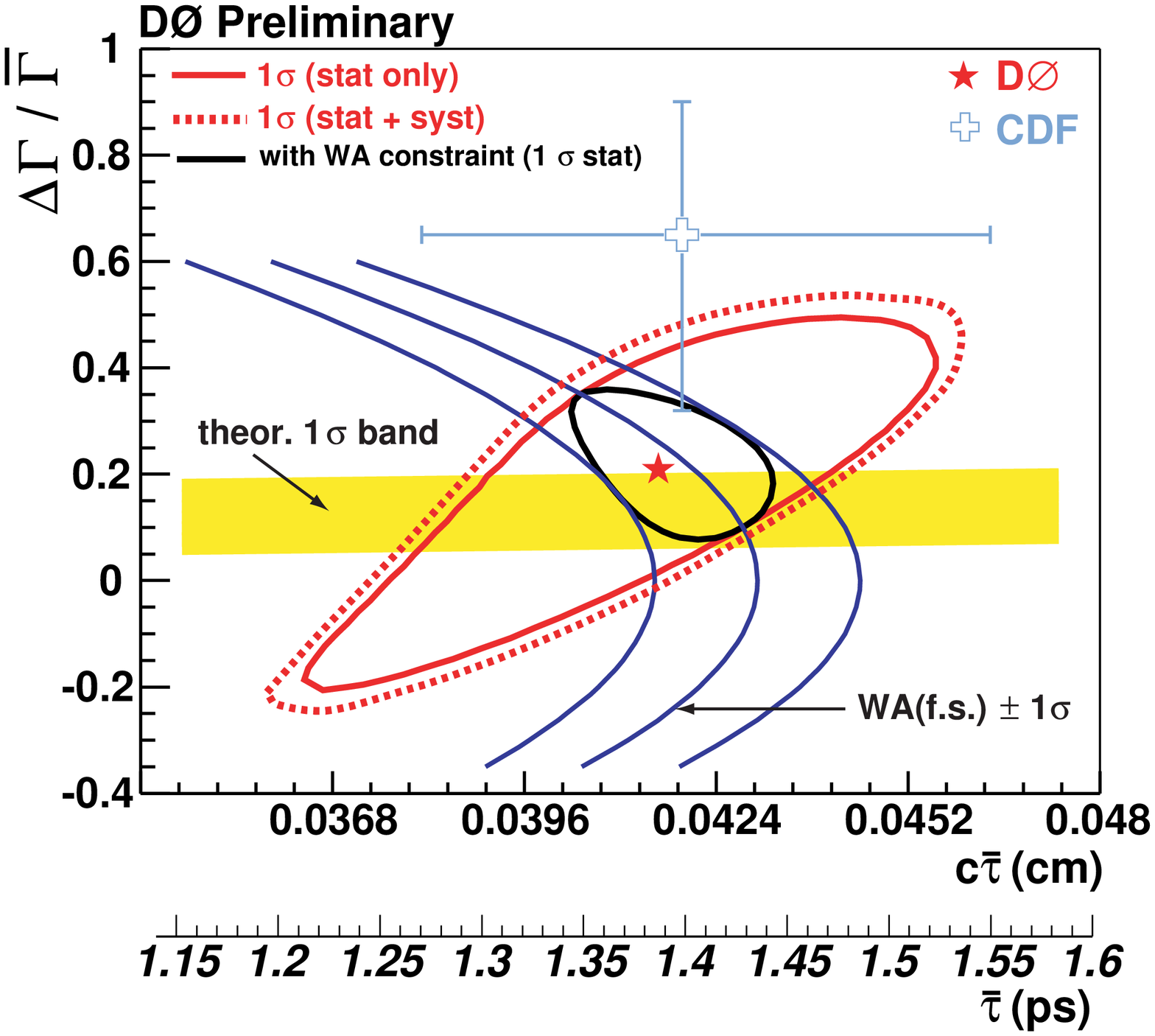}
}
\caption{$\Delta\Gamma_s/\Gamma_s$ versus c$<\tau>$ results from CDF and 
D$\oslash$.}
\label{fig:deltagammaresults}       
\end{center}
\end{figure}

The results of both experiments are summarized in Tab.~\ref{tab:deltagammaresults} 
and Fig.~\ref{fig:deltagammaresults}. The combined result slightly favors high values 
of $\Delta m_s$, but is currently statistically limited. The systematic 
uncertainties are very small, thus this is a precise measurement ones more 
data is available. 

\begin{table}
\caption{$\Delta\Gamma_s/\Gamma_s$ results from CDF and D$\oslash$.}
\label{tab:deltagammaresults}       
\begin{tabular}{lllll}
\hline\noalign{\smallskip}
  Experiment & $\Delta\Gamma_s/\Gamma_s$ & $<\tau>$ (ps) & $\tau_L$ (ps) & $\tau_H$ (ps)\\
\noalign{\smallskip}\hline\noalign{\smallskip}
 CDF        &  $0.65^{+0.25}_{-0.33}$ & $1.40^{+0.15}_{-0.13}$ & $1.05^{+0.16}_{-0.13}$  & $2.07 ^{+0.58}_{-0.46}$ \\
 D$\oslash$ &   $0.21^{+0.33}_{-0.45}$ & $1.39^{+0.15}_{-0.16}$ & $1.23^{+0.16}_{-0.13}$  & $1.52 ^{+0.39}_{-0.43}$ \\
\noalign{\smallskip}\hline
\end{tabular}
\end{table}

\section{$B_s$ Mixing}
\label{sec:deltams}

The probability that a $B$ meson decays at proper time $t$ and has or has not already mixed 
to the $\bar{B}$ state is given by:
\begin{eqnarray}
P_{unmix}(t) &\approx& \frac{1}{2}(1+\cos \Delta m t), \\
P_{mix}(t) &\approx&  \frac{1}{2}(1-\cos \Delta m t).
\end{eqnarray}

The canonical $B$ mixing analysis, in which oscillations are observed and the mixing frequency, $\Delta m$, 
is measured, proceeds as follows. The $B$ meson flavor at the time of its decay is determined by exclusive 
reconstruction of the final state. The proper time, $t=m_{B}L/pc$, at which the decay occurred is determined 
by measuring the decay length, $L$, and the $B$ momentum, $p$. Finally the production flavor must be tagged 
in order to classify the decay as being mixed or unmixed at the time of its decay.

Oscillation manifests itself in a time dependence of, for example, the mixed asymmetry:
\begin{equation}
{\cal A}_{mix}(t) = \frac{N_{mixed}(t)-N_{unmixed}(t)}{N_{mixed}(t)+N_{unmixed}(t)}=-\cos \Delta m t
\end{equation}
In practice, the production flavor will be correctly tagged with a probability $P_{tag}$, which 
is significantly smaller than one, but larger than one half (which corresponds to a random tag). 
The measured mixing asymmetry in terms of dilution, $\cal D$, is

\begin{equation}
{\cal A}^{meas}_{mix}(t) = {\cal D} {\cal A}_{mix} = -{\cal D} \cos \Delta mt
\end{equation}
where ${\cal D}=2P_{tag}-1$.

First of all a good proper decay time resolution, which is specially important in order 
to resolve high $\Delta m_s$ mixing frequency.

The second important ingredient for a mixing analysis is the flavor tagging. As the examined 
decays are flavor specific modes the decay flavor can be determined via the decay products. But 
for the production flavor additional information from the event has to be evaluated in order to tag 
the event. A good and well measured tagging performance is needed to set a limit on $\Delta m_s$.

The last component are the $B_s$ candidates. Sufficient statistic is need to be sensitive to high mixing 
frequencies.

\subsection{Flavor Tagging}

\begin{figure}
\begin{center}
\resizebox{0.50\textwidth}{0.20\textheight}{%
  \includegraphics{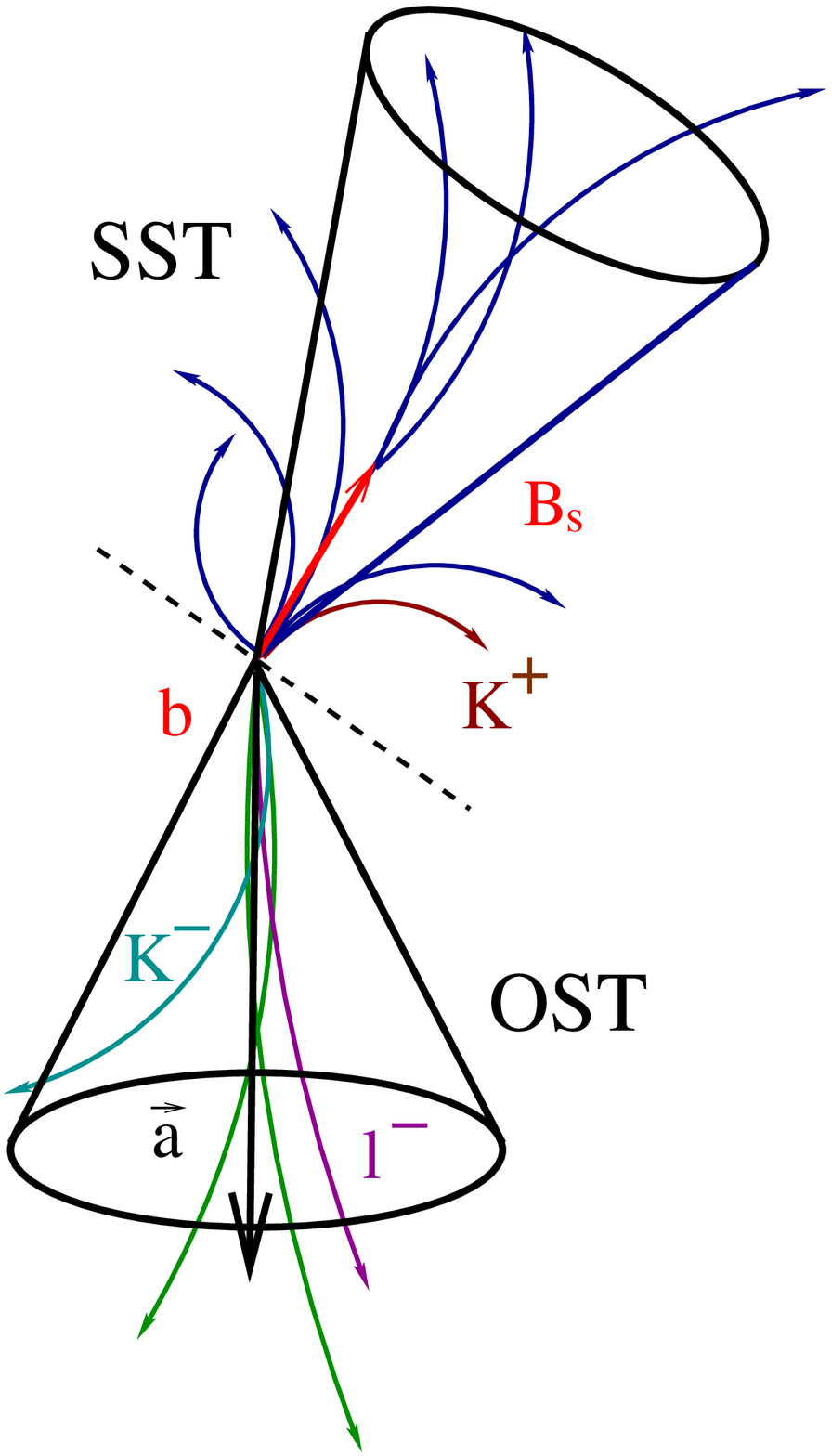}
  \includegraphics{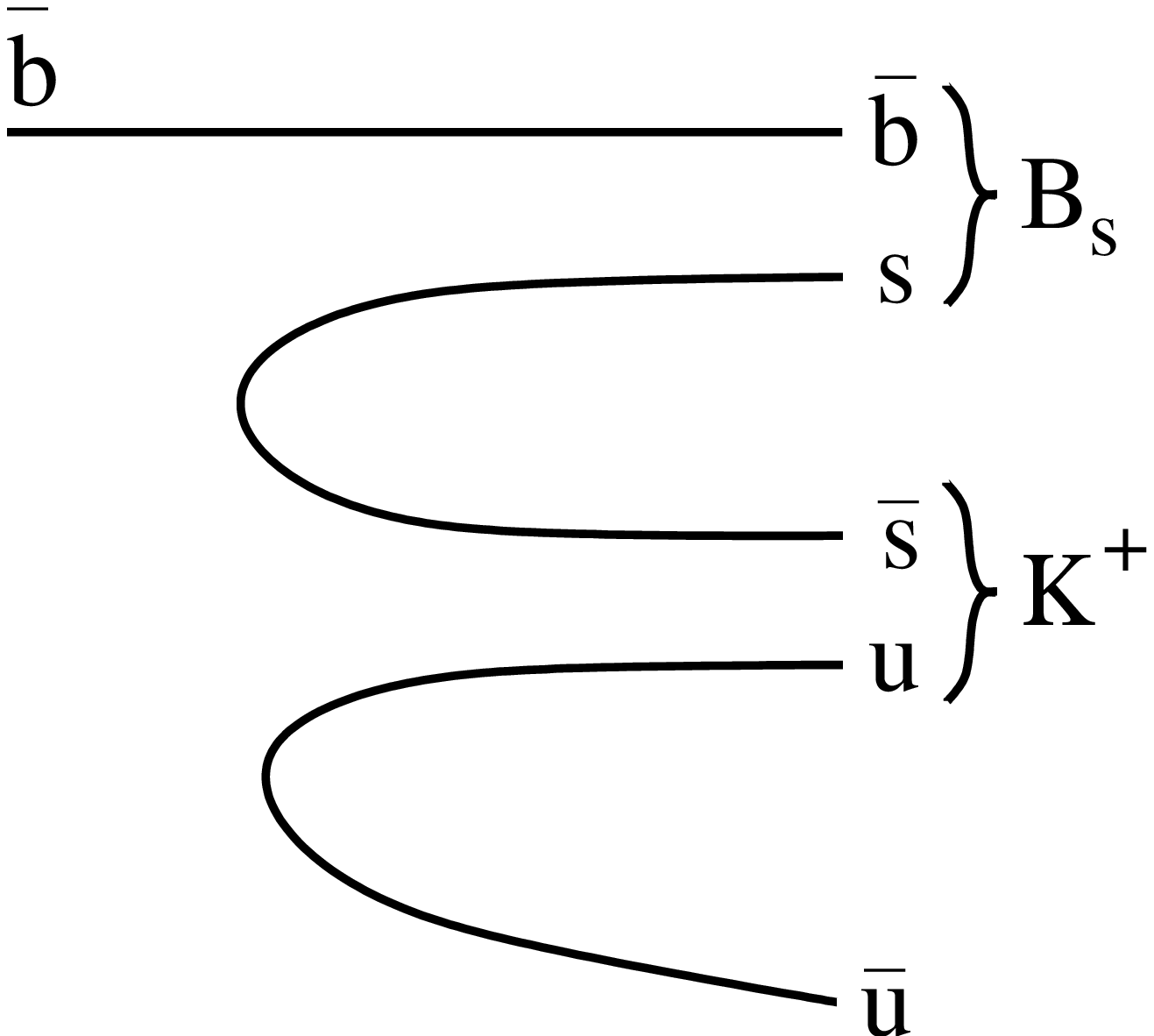}
}
\caption{Left: Sketch of different tagging algorithms; Right: Same-side kaon tagging.}
\label{fig:ceballostag1}       
\end{center}
\end{figure}

There are two different kinds of flavor tagging algorithms, opposite 
side tagging (OST) and same side tagging (SST), 
which are illustrated in Fig.~\ref{fig:ceballostag1}. OST algorithms use the 
fact that b quarks are mostly produced 
in $b\bar{b}$ pairs, therefore the flavor of the second (opposite side) $b$ 
can be used to determine the 
flavor of the $b$ quark on the signal side.

\subsubsection{Jet-Charge Tagging}
The average charge of an opposite side $b$-jet is weakly correlated to the 
charge of the opposite $b$ quark and can thus be used to determine the opposite side $b$ flavor. 
The main challenge of this tagger is
to select the $b$-jet. Information of a displaced vertex or displaced tracks 
in
the jet help to identify $b$-jets. This tagging algorithm has a very high
tagging efficiency, but the dilution is relatively low. 
By separating sets of tagged events of different qualities e.g. how $b$ like 
the jet is, it is possible to increase the overall tagging performance.

\subsubsection {Soft-Lepton-Tagging}
In 20 \% of cases the opposite semileptonic $b$ decays either into an electron 
or a muon ($b\rightarrow l^-X$). The charge 
of the lepton is correlated to the charge of the decaying $B$ meson. Depending 
on the type of the $B$ meson there is a certain 
probability of oscillation between production and decay (0 \% for $B^{\pm}$, 
17.5 \% for $B_d$ and 50 \% for $B_s$). Therefore 
this tagging algorithm already contains an intrinsic dilution. Another 
potential source of miss-tag is the transition of the $b$ 
quark into a $c$ quark, which then forms a $D$ meson and subsequently decays 
semileptonically ($\bar{b} \rightarrow \bar{c} 
\rightarrow l^-X $). Due to the different decay length and momentum 
distribution of $B$ and $D$ meson decays this source of miss-tag 
can mostly be eliminated.

\subsubsection{Same-Side-Tagging}
During fragmentation and the formation of the $B_{s/d}$ meson there is a 
left over $\bar{s}/\bar{d}$ quark which is likely to form a $K^+/\pi^+$ 
(Fig.~\ref{fig:ceballostag1}). So if there is a near by charged particle, 
which is additionally identified as a kaon/pion, it
 is quite likely that it is the leading fragmentation track and its charge 
 is then correlated to the flavor of the $B_{s/d}$ meson. While the 
 performance of the opposite side tagger does not depend on the flavor of 
 the $B$ on the signal side the SST performance depends on the signal 
 fragmentation processes.
Therefore the opposite side performance can be measured in $B_d$ mixing 
and can then be used for setting a limit on the $B_s$ mixing frequency. 
But for using the SST for a limit on $\Delta m_s$ we have to heavily 
rely on Monte Carlo simulation. The SST potentially has the best tagger 
performance, but before using it for a limit, fragmentation processes 
have to be carefully understood.

\subsection{$\Delta m_d$ Measurement and Calibration of Taggers}

For setting a limit on $\Delta m_s$ the knowledge of the tagger performance is crucial.
Therefore it has to be measured in kinematically similar $B_d$ and $B^+$ samples.

The $\Delta m_s$ and $\Delta m_d$ analysis is a complex fit with up to 500 parameters which
combine several $B$ flavor and several decay modes, various different taggers and deals with 
complex templates for mass and lifetime fits for various sources of background. Therefore the 
measurement of $\Delta m_d$ is beside the calibration of the opposite side taggers very 
important to test and trust the fitter framework, although the actual $\Delta m_d$ result at 
the Tevatron is not competitive with the $B$ factories.

Both CDF and D$\oslash$ have demonstrated that the whole machinery is working, 
being $\Delta m_d$ measurements compatible with the PDG average value of 
$0.510 \pm 0.004$ $ps^{-1}$~\cite{pdg}. The combined tagging performance of the
opposite-side taggers is about 1.5-2\%.

\begin{figure}
\begin{center}
\resizebox{0.50\textwidth}{0.20\textheight}{%
  \includegraphics{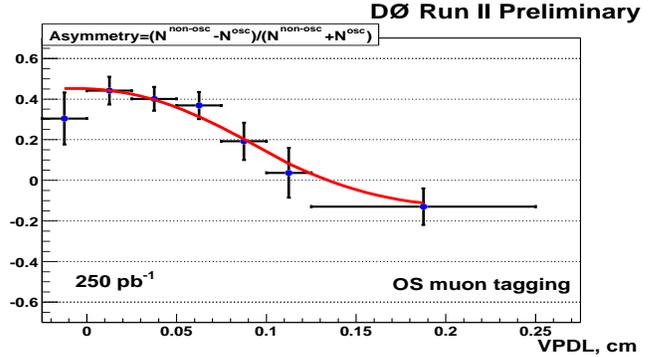}
}
\caption{Asymmetry fit projection for $\Delta m_d$ using opposite side muon tagger 
in semileptonic decays from D$\oslash$.}
\label{fig:ceballos_OST}       
\end{center}
\end{figure}

An example of the fitted asymmetry using the opposite side muon tagger on 
the semileptonic decay modes from D$\oslash$ is displayed in Fig.~\ref{fig:ceballos_OST}.  

\subsection{Amplitude Scan}
An alternative method for studying neutral $B$ meson oscillations is the so
called ``amplitude scan'', which is explained in detail in
Reference~\cite{moser}.
The likelihood term describing the tagged proper decay time of a neutral
$B$ meson is modified by including an additional parameter multiplying
the cosine, the so-called amplitude $A$.

The signal oscillation term in the likelihood of the $\Delta m$ thus becomes
\begin{equation}
{\cal L} \propto \frac{1\pm A {\cal D} \cos(\Delta m t)}{2} 
\end{equation}

The parameter $A$ is left free in the fit
while ${\cal D}$ is supposed to be known and fixed in the scan.
The method involves performing one such $A$-fit for each value of
the parameter $\Delta m$, which is fixed at each step;
in the case of infinite statistics, optimal resolution and perfect
tagger parameterization and calibration, one would expect $A$ to be
unit for the true oscillation frequency and zero for the remaining of
the probed spectrum. In practice,
the output of the procedure is accordingly a list of fitted values
($A$, $\sigma_A$) for each $\Delta m$ hypothesis.
Such a $\Delta m$ hypothesis is excluded to a $95$\% confidence
level in case the following relation is observed,
$A + 1.645 \cdot \sigma_A < 1$.

The sensitivity of a mixing measurement is defined as the lowest $\Delta m$ value 
for which $1.645 \cdot \sigma_A = 1$.

The amplitude method will be employed in the ensuing $B_s$ mixing analysis.
One of its main advantages is the fact that it allows easy combination among different
measurements and experiments.

\begin{figure}
\begin{center}
\resizebox{0.50\textwidth}{0.20\textheight}{%
  \includegraphics{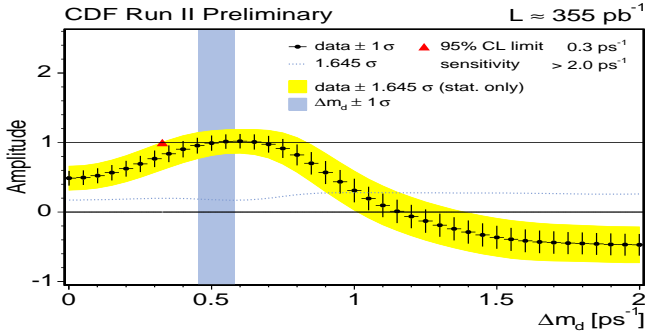}
}
\caption{Amplitude scan for $\Delta m_d$ in hadronic decay modes (CDF). 
The scan is compatible with 1 around the result of the actual 
$\Delta m_d$ fit.}
\label{fig:ceballos_amplitude}       
\end{center}
\end{figure}

The plot shown in Figure \ref{fig:ceballos_amplitude} is obtained when the 
method is applied to the hadronic $B_d$ samples of the CDF experiment, 
using the exclusively combined opposite side tagging
algorithms.

The expected compatibility of the measured amplitude with unit
in the vicinity of the true frequency, $\Delta m_d = 0.5 ~$ps$^{-1}$,
is confirmed. 

However, we observe the expected increase in the amplitude uncertainty
for higher oscillation frequency hypotheses.
This is equivalent to saying that the significance is
reduced with increasing frequency.

\subsection{Reconstructed $B_s$ Decays}

\begin{figure}
\begin{center}
\resizebox{0.50\textwidth}{0.20\textheight}{%
  \includegraphics{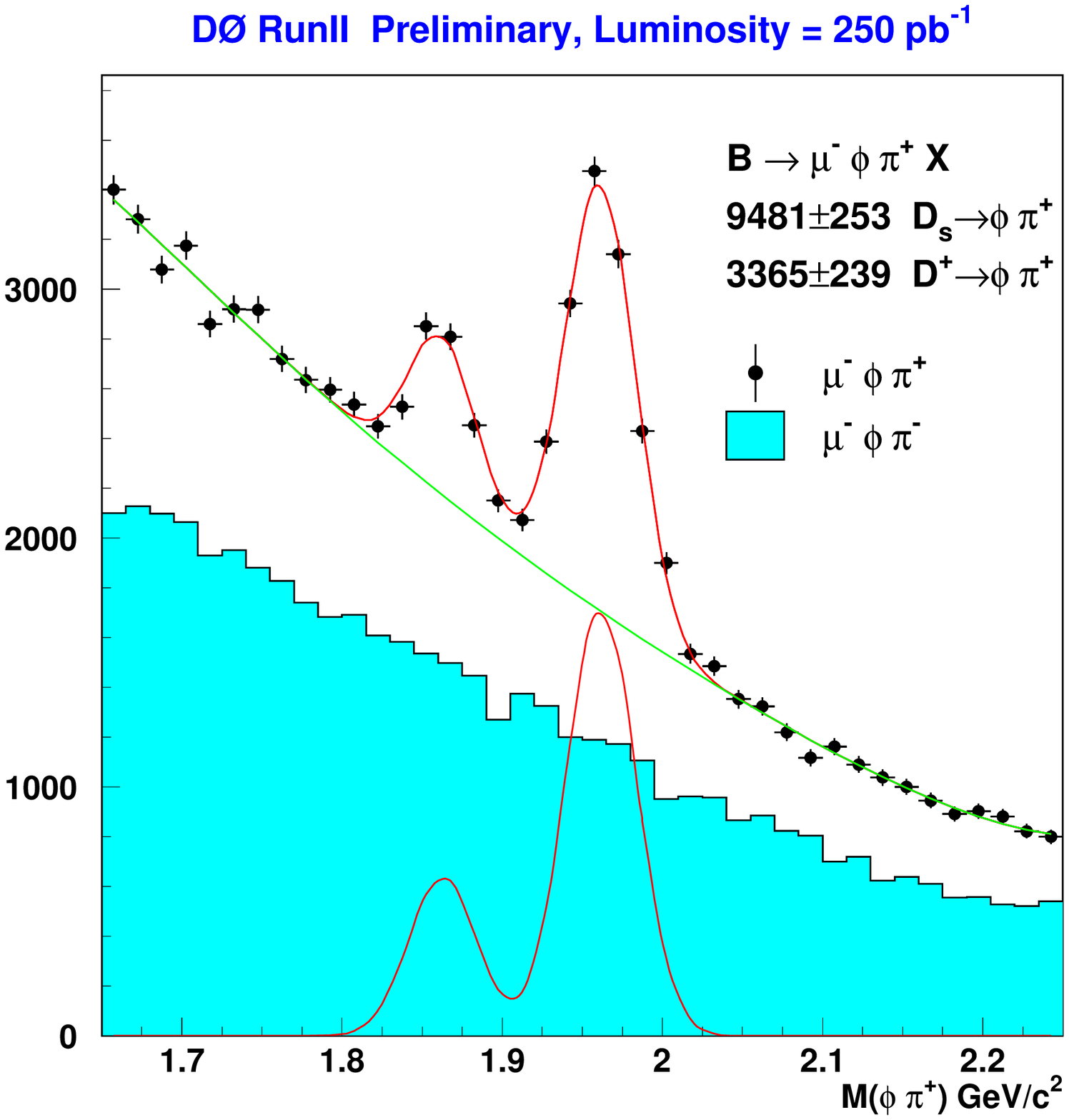}
  \includegraphics{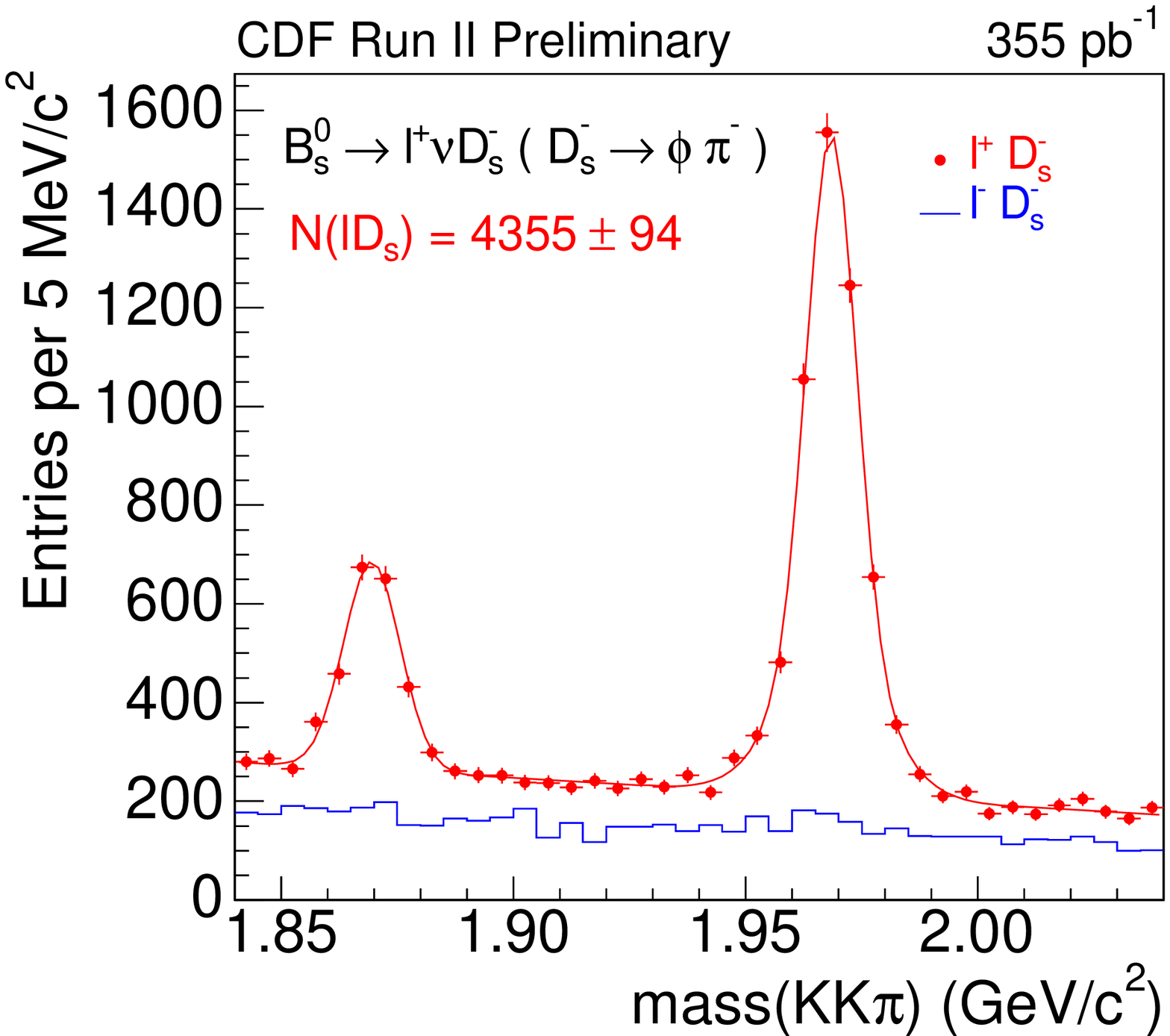}
}
\caption{Reconstructed semileptonic $B_s \rightarrow l X D_s, (D_s \rightarrow \phi \pi)$ candidates 
from D$\oslash$ (left) and CDF (right).}
\label{fig:ceballos_bsyields}       
\end{center}
\end{figure}

D$\oslash$ exploits the high statistics muon trigger to study semileptonic $B_s$ decays.
Several thousands candidates have been reconstructed in the 
$B_s \rightarrow \mu X D_s, (D_s \rightarrow \phi \pi)$ mode. Additionally D$\oslash$ is also 
working on reconstructing $B_s \rightarrow \mu X D_s, (D_s \rightarrow K^{*0}K)$ candidates and 
on reconstructing fully hadronic $B_s$ decays on the non trigger 
side in this sample.

CDF performs the $B_s$ mixing analysis using both fully reconstructed $B_s$ decays 
($B_s \rightarrow D_s \pi$) obtained by the two track trigger and semileptonic decays 
($B_s \rightarrow \ell X D_s$) collected in the lepton+displaced track trigger. In both cases 
the $D_s$ is reconstructed in the $D_s \rightarrow \phi \pi$, $D_s \rightarrow K^{*0}K$ and 
$D_s \rightarrow \pi\pi\pi$ modes.

Fig.~\ref{fig:ceballos_bsyields} shows the reconstructed semileptonic 
$B_s \rightarrow l X D_s$, $(D_s \rightarrow \phi \pi)$ candidates 
from D$\oslash$ and CDF.

\subsection{First $\Delta m_s$ Limits in Run II}

\begin{figure}
\begin{center}
\resizebox{0.50\textwidth}{0.20\textheight}{%
  \includegraphics{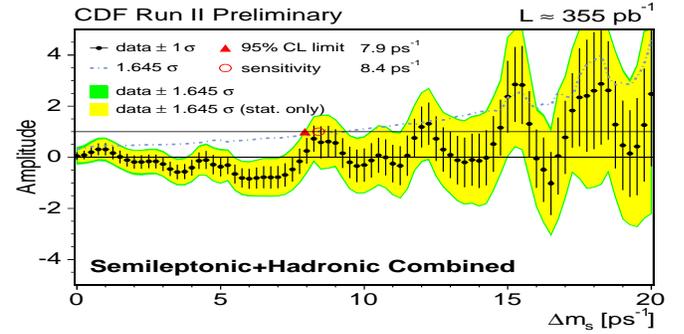}
}
\caption{Combined amplitude scan from CDF. The black dots 
represent the fitted amplitude with their respective
statistical errors for each value of $\Delta m_s$. The yellow region indicates
1.645~$\sigma$ using statistical errors only while the green band includes
combined statistical and systematic errors. 
The measurement is dominated by statistical uncertainties. Note,
neighboring points are statistically correlated.}
\label{fig:ceballos_amplitude_scan_cdf}       
\end{center}
\end{figure}

\begin{figure}
\begin{center}
\resizebox{0.50\textwidth}{0.20\textheight}{%
  \includegraphics{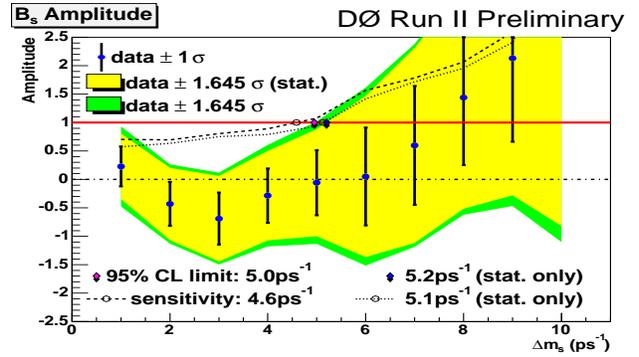}
}
\caption{Combined amplitude scan from D$\oslash$.}
\label{fig:ceballos_amplitude_scan_d0}       
\end{center}
\end{figure}

Finally, an amplitude scan, repeating the Likelihood fit for the amplitude 
$A$ for different values of $\Delta m_s$, was performed in both D$\oslash$ 
and CDF. The results of the amplitude scans are shown in 
Fig.~\ref{fig:ceballos_amplitude_scan_cdf} 
and~\ref{fig:ceballos_amplitude_scan_d0}. 
The amplitude scan yields a $\Delta m_s$~sensitivity of 8.4(4.6)~$ps^{-1}$ 
and a lower exclusion limit of 7.9(5.0)~$ps^{-1}$ is set on the  value of 
$\Delta m_s$ at a 95\% confidence level in CDF (D$\oslash$). 

Those results are good enough for the first round of the analysis, but there is
still a huge room for improvements in the near future.

\section{Conclusions}
\label{sec:conclusions}
The large amount of data collected by the CDF and D$\oslash$ experiments are improving 
our knowledge about $B_s$ mesons. A few selected topics have been discussed in this 
paper. The measurement of the decay width difference $\Delta 
\Gamma_s$ of the heavy and light $B_s$ mass eigenstate is especially sensitive to 
high $\Delta m_s$ values. The $B_s$ mixing analysis is sensitive to lower values. 
Together they have the potential to cover the hole range of possible $\Delta m_s$ 
values in the Standard Model and as well beyond.

\end{document}